\documentclass[journal]{IEEEtran}
\usepackage{citesort}
\usepackage{amsmath}
\usepackage{epsfig}
\usepackage{subfigure}
\usepackage{enumerate}
\usepackage{setspace}
\usepackage{url}

\newcommand\blfootnote[1]{%
  \begingroup
  \renewcommand\thefootnote{}\footnote{#1}%
  \addtocounter{footnote}{-1}%
  \endgroup
}

\begin{document}
\title{A Blind Zone Alert System based on\\ Intra-vehicular Wireless Sensor Networks}
\author{\authorblockN{Jiun-Ren Lin, \textit{Student Member}, \textit{IEEE}, Timothy Talty, \textit{Senior Member}, \textit{IEEE}, and Ozan K. Tonguz, \textit{Member}, \textit{IEEE}}}
\maketitle

\blfootnote{J.-R. Lin and O. K. Tonguz are with the Department of Electrical and Computer Engineering, Carnegie Mellon University, Pittsburgh, PA 15213, USA (email: j.lin.us@ieee.org; tonguz@ece.cmu.edu)}
\blfootnote{Timothy Talty is with General Motors Company, Detroit, MI 48232, USA (email: timothy.talty@gm.com)} 

\blfootnote{Copyright \textcopyright 2015 IEEE. Personal use of this material is permitted. However, permission to use this material for any other purposes must be obtained from the IEEE by sending a request to pubs-permissions@ieee.org}

\begin{abstract}
Due to the increasing number of sensors deployed in modern vehicles, Intra-Vehicular Wireless Sensor Networks (IVWSNs) have recently received a lot of attention in the automotive industry as they can reduce the amount of wiring harness inside a vehicle. By removing the wires, car manufacturers can reduce the weight of a vehicle and improve engine performance, fuel economy, and reliability. In addition to these direct benefits, an IVWSN is a versatile platform that can support other vehicular applications as well. An example application, known as a Side Blind Zone Alert (SBZA) system, which monitors the blind zone of the vehicle and alerts the driver in a timely manner to prevent collisions, is discussed in this paper. The performance of the IVWSN-based SBZA system is evaluated via real experiments conducted on two test vehicles. Our results show that the proposed system can achieve approximately $95\%$ to $99\%$ detection rate with less than $15\%$ false alarm rate. Compared to commercial systems using radars or cameras, the main benefit of the IVWSN-based SBZA is substantially lower cost.
\end{abstract}

\begin{IEEEkeywords}
wireless sensor networks, vehicle safety, automotive sensors, blind zone detection, Bluetooth Low Energy, vehicular networks
\end{IEEEkeywords}

\IEEEpeerreviewmaketitle

\section{introduction}
Modern production vehicles are highly computerized, and the major functionalities of a vehicle are controlled by several Electronic Control Units (ECUs) inside the vehicle. ECUs require sensors to gather real-time information of the vehicle in order to control the vehicular operations. Currently, most of the sensors inside vehicles are connected to the ECUs by physical wires. Controller Area Networks (CAN), FlexRay, and Local Interconnect Network (LIN) are the common technologies currently used for the wired network inside vehicles. However, because the complexity of vehicles is getting higher, and the number of applications and gadgets in vehicles keeps increasing, the large number of wires needed to connect sensors with ECUs pose significant challenges: i) wires bring extra weight to the vehicle; ii) wiring limits locations where sensors can be installed; iii) cost of the wires and wiring may be high; and iv) wires can be worn down with time and locating and/or replacing such wires can be very costly and cumbersome. Therefore, reducing the wires in the vehicle could help enhance the fuel economy, performance, product features, and reliability as well as reduce the overall cost of building a vehicle. To achieve these goals, wireless technology can be applied to the communications between sensors and ECUs and replace the wired connections. In the remainder of this paper, the new platform is called an Intra-Vehicular Wireless Sensor Network (IVWSN)~\cite{GM-intra-vehicular}.

Over the last few years, researchers and engineers in wireless and automotive industries have been working on various domains related to Wireless Sensor Networks (WSNs)~\cite{yick2008wireless}, including architecture~\cite{Kong2008}, Medium Access Control (MAC) protocols~\cite{demirkol2006mac}\cite{ergen2010tdma}\cite{ergen2006pedamacs}\cite{shen2014prioritymac}\cite{toscano2012multichannel}, routing protocols~\cite{akkaya2005survey}\cite{ergen2007energy}\cite{zhang2014energy}\cite{niu2014r3e}, system design~\cite{bonivento2007system}, application design~\cite{shin2010experimental}, energy conservation~\cite{anastasi2009energy}\cite{anastasi2009extending}\cite{magno2014ensuring}, and quality of service~\cite{li2014qos}\cite{marchenko2014experimental}\cite{park2014robust}. Although the basic concept of IVWSNs stems from classical WSNs, IVWSNs have several unique characteristics and require special consideration in terms of system and protocol design. For example, although the sensors and ECUs inside a vehicle are either fixed or have predictable movements within a small area, there are metal parts, obstacles, and passengers inside one vehicle, and it creates an especially challenging environment for radio propagation. Furthermore, many vehicular applications have stringent requirements in terms of the reliability and latency of the communications. All of these special characteristics of IVWSNs have to be taken into account for the system design (including the protocol design) of an IVWSN. Existing research works mainly focused on channel assessment~\cite{intra-car-channel}, evaluation of wireless technologies for IVWSNs~\cite{Lin_Globecom2013}, and security issues~\cite{lee2009security}. Regarding the wireless technologies for IVWSNs, various investigations have been conducted on ZigBee~\cite{Tsai07}, Bluetooth Low Energy (BLE)~\cite{BLELin14}, radio-frequency identification (RFID)~\cite{Tsai06}, ultra-wideband (UWB)~\cite{niu2008intra}\cite{bas2012ultra}\cite{bas2013ultra}, 2.4 GHz customized RF~\cite{carmo20102}, and coexistence of multiple wireless technologies~\cite{Lin_Globecom2013}\cite{de2009coexistence}\cite{lo2009coexistence}. In addition to passenger vehicles, similar ideas have been applied to aircraft as well~\cite{matolak2008aircraft}. The main focus of this paper is the potential industrial applications of IVWSNs. 

\begin{figure}[tbp]
\centering
\psfig{figure=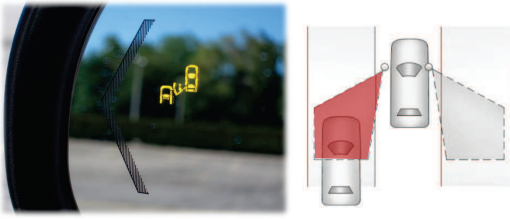,width=0.38\textwidth}
\caption{General Motors' SBZA system on a 2011 Cadillac Escalade~\cite{2011EscaladeManual}}
\label{BZA}
\end{figure}

One of the industrial safety features currently implemented in some modern vehicles is the side blind zone alert (SBZA) system, which monitors the blind zone of the vehicle and alerts the driver accordingly in a timely manner to prevent collisions.\footnote{According to National Highway Traffic Safety Administration (NHTSA), lane change is the fourth U.S. leading cause of crashes. It contributes to $9\%$ of all crashes; other causes include rear-end ($28\%$), crossing path ($25\%$), and Run-off-road ($23\%$)~\cite{NHTSA_Plan_2013}.} Fig.~\ref{BZA} illustrates the SBZA system from General Motors (GM) on a 2011 Cadillac Escalade. However, since most of the existing SBZA systems are based on radar, it is difficult to lower the cost of the system and deploy in lower-end vehicles. Also, it has been shown that a radar-based system could suffer from false alarms, which are triggered by tree leaves, pedestrians, and other nearby objects. Motivated by the aforementioned two limitations of the radar-based systems, this paper proposes to use an IVWSN to support the blind zone alert feature as a new industrial platform.

Although an IVWSN is designed for communications within a vehicle, some of the sensor packets could reach other vehicles at a close distance, e.g., less than 10 meters, depending on the transmission power used. Therefore, the underlying concepts and the IVWSN platform could be utilized for inter-vehicular communications within a small range. Moreover, some of the vehicular sensors may have a low duty cycle, and hence those sensors are idle during most of the time. Those sensors could be reused during the idle time for other applications in order to achieve better utilization and lower overall cost. For example, since 2006, it is mandated by law in the United States that all vehicles manufactured in 2006 and onward must be equipped with the Tire Pressure Monitoring System (TPMS)~\cite{TPMS_NHTSA}. Since wireless sensors of the TPMS have a low duty cycle (e.g., 0.01\% to 5\%), in principle, one could use such sensors to broadcast beacon packets for blind zone detection on adjacent vehicles. As a proof of concept, in this paper, a blind zone alert system is designed and implemented on an IVWSN platform based on the BLE technology~\cite{BLEspec}. BLE is a wireless technology introduced in 2010 along with Bluetooth specification version 4.0 and could be a great fit for certain applications of IVWSNs due to its low-power, low-complexity, and low-cost properties~\cite{BLELin14}. It is worth pointing out that in addition to BLE, the proposed system could be also implemented using other wireless technologies with minimal modifications, such as ZigBee or other proprietary low-power wireless technologies. The detection algorithm proposed in this paper is based on a Neyman-Pearson (NP) classifier. Its simplicity can help lower the cost of the system. It is also flexible since the NP-threshold parameter can be adjusted to satisfy the desirable detection and false alarm rates.

This paper is organized as follows. Section II describes the details of the experimental setup used and the blind zone detection algorithm. Section III presents the experimental results and the performance of the proposed system. Section IV discusses the major issues related to the proposed design. Finally, concluding remarks are given in Section V.

\section{Experimental study}
\subsection{Experimental platform}
\begin{figure}[tbp]
\centering
\subfigure[Texas Instruments CC2540 Mini Development Kit]{
\includegraphics[scale = 0.28]{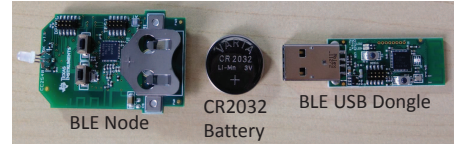}
\label{platform_fig:subfig1}
}
\subfigure[System architecture]{
\includegraphics[scale = 0.38]{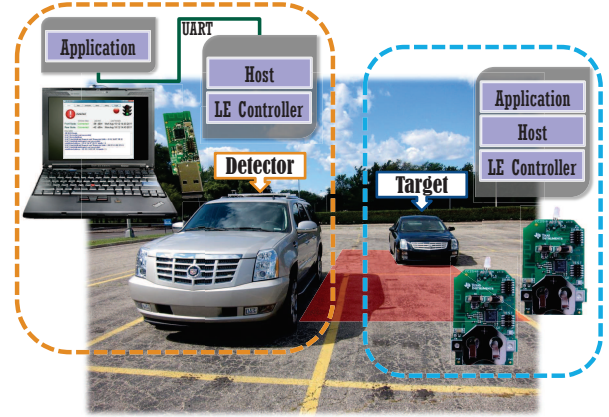}
\label{platform_fig:subfig2}
}
\caption[]{Bluetooth Low Energy experimental platform}
\label{BLE_Stack}
\end{figure}

The IVWSN platform in this study is developed on the Texas Instruments CC2540 Mini Development Kit~\cite{CC2540kit}. Texas Instruments CC2540 is a single-chip industrial BLE solution which can run the BLE protocol stack and applications with a built-in 8051 microcontroller. The development kit comes with a BLE node and a BLE USB Dongle, as shown in Fig.~\ref{platform_fig:subfig1}. The default power source of the BLE node is a CR2032 coin battery. The setup of the experimental platform is shown in Fig.~\ref{platform_fig:subfig2}. The USB dongle is connected to a personal computer (PC) with a USB to serial link. The PC emulates the functionalities of an ECU in the real automotive platform. On the USB dongle, there are the Host, an LE Controller, and an adaptation layer, which serves as the interface between the Host and the PC. The application layer, where the blind zone detection algorithm resides, and a serial port interface are implemented on the PC. The USB dongle along with the PC is the detection device of the proposed blind zone alert system, while the BLE node represents the wireless sensor that broadcasts beacons. On the BLE node, there are the application layer, the Host, and an LE Controller. Note that the Host and the LE Controller are parts of the BLE standard protocol stack.

\begin{figure}[tbp]
\centering
\includegraphics[scale = 0.42, trim = 1mm 1mm 0mm 1mm, clip]{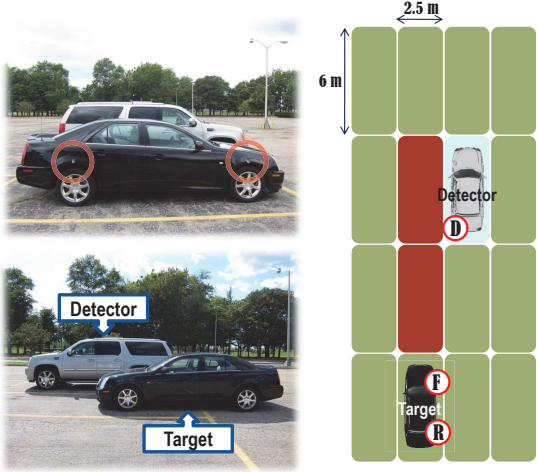}
\caption{Setup of Experiment 1 (parking scenario)}
\label{SBZAlot}
\end{figure}

As shown in Fig.~\ref{SBZAlot}, a 2008 Cadillac Escalade is the detector vehicle, which tries to determine if the target vehicle (i.e., a 2008 Cadillac STS) is within its blind zone. The detection device is installed on the left side of the rear bumper of the detector vehicle while two BLE nodes (i.e., the sensors) are installed right above the two right wheels of the target vehicle (see Fig.~\ref{SBZAlot}). The two BLE nodes represent the sensors on the front and the rear wheels, respectively, and they are referred to as the front sensor and the rear sensor. The detection device continuously monitors the Received Signal Strength Indicator (RSSI) of beacon packets broadcasted by the two sensors on the target vehicle. 

In the empirical evaluation process, each sensor on the target vehicle sends out four beacon packets every second. Note that in practice, existing wireless sensors such as TPMS sensors can be used to periodically send out the packets, so it is not necessary to install new dedicated sensors for this application. On the detector vehicle, the detection device measures the RSSI of the received beacon packets, and in order to have a better ability to distinguish the target vehicle at different positions, the detection device utilizes the RSSI values of beacon packets from both the front and the rear sensors.
An observation collected by the detection device, denoted by $x_t$, can be represented by
\begin{align}
x_t &\doteq ({RSSI}_{front}, {RSSI}_{rear})
\end{align}
where ${RSSI}_{front}$ is the RSSI of the received beacon packet, which originates from the front sensor, and ${RSSI}_{rear}$ is the RSSI of the packet from the rear sensor. 
Also, in some circumstances, there might be multiple cars in the detector vehicle's communications range. The detection device has to combine the two corresponding RSSI values of packets from the two sensors on the same vehicle as one observation record. Therefore, the type/model/identification information of the vehicle and the sensor location (i.e., front or rear; left or right) are included in the header of beacon packets. In the evaluation process, each observation is associated with an actual detection state (i.e., whether or not a target is actually in the detector's blind zone). These pieces of information are then stored in the PC and are later used in both the training and testing phases (which is the standard method for many machine learning algorithms); the training phase is performed to train the decision mechanism and obtain the decision map while the testing phase is used to evaluate the accuracy of the system.

\subsection{Blind zone detection mechanism}
\begin{figure}[tbp]
\centering
\includegraphics[scale = 0.42, trim = 0mm 0mm 0mm 0mm, clip]{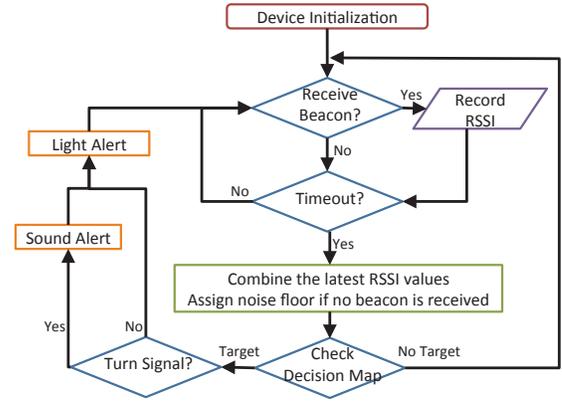}
\caption{Flowchart of the SBZA based on IVWSNs}
\label{BZA_flow}
\end{figure}

Fig.~\ref{BZA_flow} is the flowchart that illustrates the principle of operation of the detection device on the detector vehicle. The detection device periodically collects the beacon packets from the sensors on the target vehicle. At the end of each period, the detection device combine the RSSI values of the received packets from the sensors and compute the moving weighted average of the last three observations.\footnote{This is done to handle the fading phenomena of the wireless channel.} Therefore, the average observation $x_i'$ can be derived from the equation
\begin{align}
x_i' &\doteq (x_{nT} + x_{(n-1)T} + x_{(n-2)T})/3
\end{align}
where $x_{nT}$ is the current observation, $x_{(n-1)T}$ is the previous one, and so on; $i$ is the index of the observations made in the training phase. If no beacon packet is received during the period, an RSSI value equal to the noise floor will be assigned for that observation. According to the average observation $x_i'$ and the decision map (obtained during the training phase; details are given in the following subsection), the detector determines if the target is in its blind zone. If the target is determined to be in the blind zone (i.e., case \emph{Target}) and the driver is using the turn signal (which means the potential collision is imminent), the system will use both sound and light alerts to notify the driver effectively; if the driver is not using the turn signal at the moment, the system will just use the light alert.

\subsection{Experimental setup}
Two experiments were performed to evaluate the proposed blind zone alert system. The first experiment was done in an empty parking lot in GM Technical Center, in Warren, MI, USA. As shown in Fig.~\ref{SBZAlot}, the detector vehicle was parked in the middle of the parking lot, whereas the target vehicle moves forward or backward in the lanes around the detector vehicle (e.g., it may be located in the green area or the red area around the detector vehicle as shown in the figure). The speed of the target vehicle varied between 0 and 20 mph.
The red area represents the blind zone of the detector vehicle.\footnote{Note that the red area includes guard spaces so it is larger than the actual blind zone.} The target vehicle is considered in the blind zone of the detector vehicle if it is located in the red area, and $C_1$ represents this class (i.e., case \emph{Target}). The other class is $C_2$, which means that there is no target vehicle in the blind zone of the detector vehicle (i.e., case \emph{No Target}). During the training phase, the observations $x_i'$ along with the actual class are recorded. Therefore, each record $r_i$ can be represented as
\begin{align}
r_i &\doteq (x_i',g_i)
\end{align}
where $g_i$ is the actual class of the target vehicle, which can be either $C_1$ or $C_2$. Two random variables, i.e., $X$ and $G$, represent the observations and the actual class recorded during the training phase, respectively. The durations of the experiments are 60 minutes for the training phase and 30 minutes for the testing phase. A total of 14400 records ($r_i$) were gathered for the training phase and the obtained decision map was used to evaluate the additional 7200 observations.

\begin{figure}[tbp]
\centering
\psfig{figure=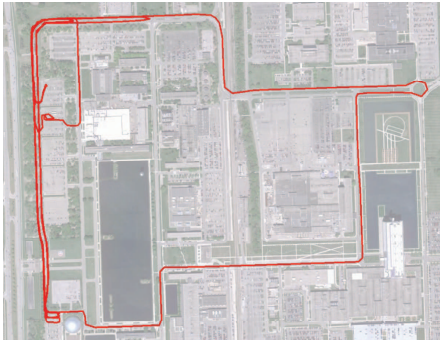,width=0.32\textwidth}
\caption{GPS trip log of Experiment 2 (driving scenario)}
\label{GPSMAP}
\end{figure}

Contrary to the parking scenario, the second experiment was performed when both cars were moving. Both vehicles were driven along a 10-mile track in the GM Technical Center in Warren, MI, USA (see Fig.~\ref{GPSMAP}); during the experiment, the target vehicle moved in and out of the detector's blind zone. The durations of the experiments are 60 minutes for the training phase and 30 minutes for the testing phase. The speeds of both vehicles varied between 0 and 40 mph. A total of 10800 and 4800 records were collected during the training and testing phases, respectively. Note that due to the dynamics of locations of the vehicles, the GM's commercial radar-based SBZA system is used as the ground truth in the experiment (i.e., it is used to determine if the target vehicle is {\it actually} in the blind zone). A 2011 GMC Yukon and a 2008 Cadillac STS are used as the detector and target vehicles, respectively.

\subsection{Determining the decision map}
\begin{figure}[tbp] %
\centering
\subfigure[Class \emph{Target}: $P(X|G=C_1)$]{
\includegraphics[scale = 0.48, trim = 15mm 5mm 15mm 17mm, clip]{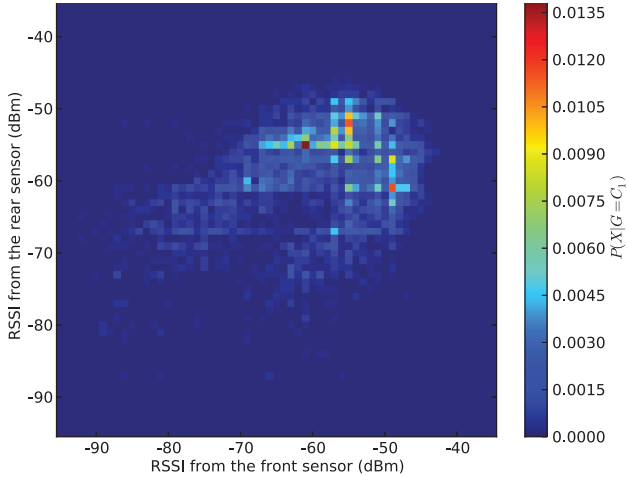}
\label{fig:subfig1}
}
\subfigure[Class \emph{No Target}: $P(X|G=C_2)$]{
\includegraphics[scale = 0.48, trim = 15mm 5mm 15mm 17mm, clip]{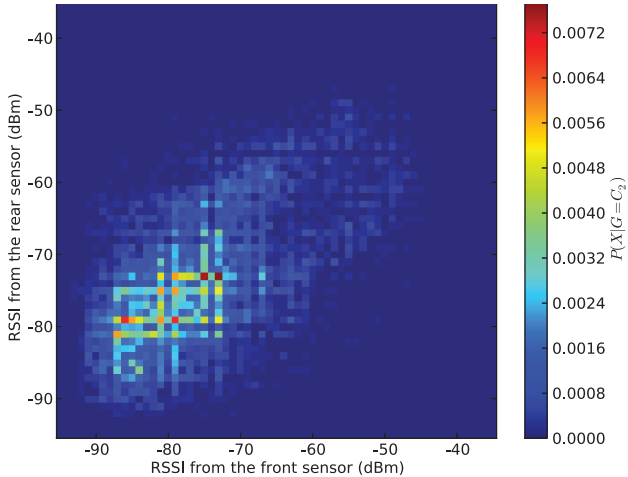}
\label{fig:subfig2}
}
\caption[]{Normalized histograms of the observations collected during training phase
}
\label{BZAS_Histogram}
\end{figure}

\begin{figure}[htp]
\centering
\includegraphics[scale = 0.48, trim = 10mm 5mm 11mm 15mm, clip]{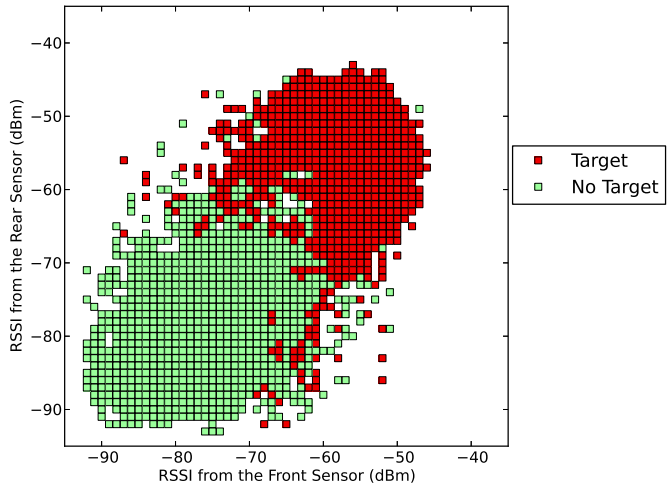}
\caption{Decision map of Experiment 2 when $\lambda=0.5$}
\label{DMap}
\end{figure}

The blind zone detection problem considered here can be formulated as a binary classification problem; the \emph{Target} class ($C_1$) represents the case in which the target vehicle is in the blind zone of the detector vehicle and the \emph{No Target} class ($C_2$) otherwise. In this paper, because of its flexibility and low computation complexity, a NP classifier, which is commonly used in radar research, is chosen. Briefly, the decision rule of the NP classifier can be stated as follows:
\begin{align}
d_j &\doteq  C_1\mbox{, if } \frac{P(X=x_j'|G=C_1)}{P(X=x_j'|G=C_2)} > \lambda\mbox{;}\\
d_j &\doteq  C_2\mbox{, if } \frac{P(X=x_j'|G=C_1)}{P(X=x_j'|G=C_2)} < \lambda\mbox{}
\end{align}
where $\lambda$ is the NP user-defined threshold. Also, for the boundary case,
\begin{align}
d_j &\doteq C_1 \mbox{, if } \frac{P(X=x_j'|G=C_1)}{P(X=x_j'|G=C_2)} = \lambda\mbox{ and } Y=0\mbox{;}\\
d_j &\doteq C_2 \mbox{, if } \frac{P(X=x_j'|G=C_1)}{P(X=x_j'|G=C_2)} = \lambda\mbox{ and } Y=1\mbox{}
\end{align}
where $d_j$ is the decision according to the observation $x_j'$, and $j$ is the index of the observations made in the testing phase. Y is a random variable following a Bernoulli distribution with a predefined probability $p$. $P(X=x_j'|G=C_1)$ is the conditional probability of having an observation $x_j'$ given that there is a car in the blind zone, and $P(X|G=C_1)$ can be visualized as a 2-D (two-dimensional) histogram. Based on the set of the training record $r_i$, $P(X=x_j'|G=C_1)$ and $P(X=x_j'|G=C_2)$ can be calculated through the following equations:
\begin{align}
P(X=x_j'|G=C_1) &=\frac{P(X=x_j', G=C_1)}{P(G=C_1)};\\
P(X=x_j'|G=C_2) &=\frac{P(X=x_j', G=C_2)}{P(G=C_2)}.
\end{align}
Fig.~\ref{fig:subfig1} and Fig.~\ref{fig:subfig2} represent $P(X|G=C_1)$ and $P(X|G=C_2)$, respectively. Once the threshold $\lambda$ is assigned, a decision map (as shown in Fig.~\ref{DMap}) can be determined by applying the aforementioned decision rule to all the possible observations. 

The performance metrics of the detection algorithm are the detection rate (i.e., $P_D$) and the false alarm rate (i.e., $P_{FA}$), and they can be defined as
\begin{align}
P_D &= P(d_j = C_1 | g_j = C_1);\\
P_{FA} &= P(d_j = C_1 | g_j = C_2).
\end{align}
The probability is calculated based on all the decision $d_j$ and actual class $g_j$ of the observation collected during the testing phase.
$P_D$ is the probability that the algorithm correctly classifies the observation into the \emph{Target} class when the observation actually belongs to the \emph{Target} class; $P_{FA}$ is the probability that the algorithm misclassifies the observation into the \emph{Target} class when the observation actually belongs to the \emph{No Target} class. Furthermore, it is worth pointing out that the threshold $\lambda$ is an adjustable parameter: if $\lambda$ takes on a small value (i.e., close to 0), major portion of the decision will fall into the target class $C_1$ and vice versa. In other words, with a close-to-zero $\lambda$ value, the algorithm almost always classifies the observations into the target class, resulting in a high detection rate. However, it must be noted that although the high detection rate is desirable, small $\lambda$ also leads to a high false alarm rate, which is not desirable. As a result, the goal of the system is to maximize the detection rate while keeping the false alarm rate under a certain tolerable level. By varying the threshold from $0$ to $\infty$, different operating points $(P_{FA},P_D)$ of the system can be achieved (see the Receiver Operating Characteristic (ROC) curves in Fig.~\ref{ROC_both}).

\section{Experimental Results}
\begin{figure}[tbp]
\centering
\includegraphics[scale = 0.47, trim = 10mm 5mm 15mm 9mm, clip]{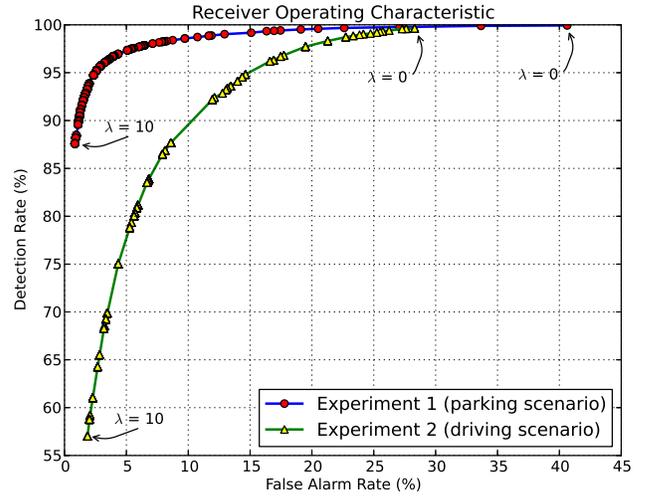}
\caption{ROC of the classifier}
\label{ROC_both}
\end{figure}

Results of both experiments are shown in Fig.~\ref{ROC_both} at which the values of lambda varies from 0 to 10 with an increment of 0.01. Observe that as the value of lambda decreases (e.g., from 10 to 0), both the false alarm rate and the detection rate increase. For instance, in the first experiment, the detection rate and false alarm rate increase from 97.34\% to 99.2\% and from 5.08\% to 19.1\%, respectively, as the lambda decreases from 0.26 to 0.005. Similar observations can be made for Experiment 2.

In addition, observe that the performance of the blind zone detection system is worse when one considers a driving scenario (i.e., Experiment 2). For example, with approximately 19\% false alarm rate, one could obtain 99.2\% detection rate in the parking scenario whereas only 97.72\% can be achieved in the driving scenario. The performance degradation in the driving scenario is a result of two fundamental reasons: the fading effect of the wireless channels when vehicles are moving and the artifact created by the GM's commercial SBZA system (more details are given in Section IV).

According to the study conducted, the proposed system based on the NP classifier leads to a comparable performance to an alternative design that utilizes a K Nearest Neighbor (KNN) classifier~\cite{JRVTC2011}. However, the alternative system based on KNN is not as flexible as the proposed system since it is not able to make the trade-off between detection rate and false alarm rate. Furthermore, the NP classifier does not require as much computational power, which could be significant when the number of observations is large. It is also worth noting that the decision map (as shown in Fig.~\ref{DMap}) in the proposed system can be precomputed during the training phase, and hence the detection becomes an efficient table look-up operation. Also, the detection device just needs to store the decision map in the memory, so the memory requirement will be minimum as well. Because of the low computational power and memory requirements, it is possible to implement the proposed system on a low-cost embedded system.

\begin{table}[tbp] %
	\centering
		\caption{The detection rate of the BLE-based Blind Zone Alert System}
		\resizebox{0.36\textwidth}{!}{
		\begin{tabular}{|c||c|c|}
			\hline 
		 	\textbf{Threshold ($\lambda$)} & \textbf{Detection Rate} & \textbf{False Alarm Rate}\\
		 	\hline \hline
		 	\multicolumn{3}{|c|}{Experiment 1 (parking scenario)}\\
		 	\hline
		 	0.26 & 97.34\% & 5.08\%\\ \hline
		 	0.1 & 99.03\% & 12.94\% \\ \hline
		 	0.005 & 99.2\% & 19.1\% \\ \hline\hline
		 	\multicolumn{3}{|c|}{Experiment 2 (driving scenario)}\\
		  \hline
		  0.71 & 94.81\% & 14.6\%\\ \hline
		  0.7 & 96.19\% & 16.6\%\\ \hline
		  0.36 & 97.72\% & 19.48\%\\ \hline
		\end{tabular}
		}
	\label{DetectionRate}
\end{table}

\section{Discussion}
\subsection{Interpretation of Results}
Based on the experimental results reported in the previous section, it is observed that the detection and false alarm rates of the system highly depend on the experimental setup and the definition of the blind zone. It has been observed that most of the missed detections and false alarms occur when the target vehicle is at the boundary of the specified blind zone (i.e., it either just enters or is about to leave the blind zone). Since under normal driving conditions, it is not likely that the target vehicle will stay at the edge of the blind zone, the actual detection rate of the system (i.e., observed in the normal condition) can be potentially higher than the results reported here. Furthermore, since the proposed system only performs detection when valid packets are received, it reduces the potential false alarms caused by other objects near or in the blind zone. In the reported experimental results, the false alarm rates were calculated based on only the observations $x_t$ gathered during the experiments. However, during the time that the system didn't receive any packet, the system should be considered as making correct dismissals. Therefore, in the real use cases, the false alarm rate that the user experiences is actually smaller.  

\subsection{Comparison to systems based on radars or cameras}
Radar is considered to be a reliable tool for detecting objects and obstacles, and it has been used for vehicle detection on commercial systems for many years. However, in general, radar is an expensive instrument, and its high cost prevents it from being massively deployed in passenger vehicles. Therefore, an alternative design with lower-cost is very desirable. Furthermore, since a radar could not tell the difference between vehicles and other objects, a blind zone alert system based on radars could suffer from false alarms due to pedestrians, Jersey barriers, leaves, rain and snow, and other objects in the detection range~\cite{2011EscaladeManual}\cite{RITABSWlesson}. On the other hand, the proposed system detects the beacon packets from other vehicles, so it could have a much lower false alarm rate.

It is also possible to use a camera to detect the vehicle in the blind zone~\cite{sotelo2007vision}\cite{alonso2008lane}. However, the performance of vision-based detection on vehicles highly depends on the visibility, light and weather conditions of the environment. Furthermore, slight difference in the relative position between vehicles could cause miss detections or false alarms on a vision-based system, and even the curvature of the road could affected its performance as well~\cite{Lee2013nhtsa}. Nevertheless, there is currently no standardized method to evaluate the detection and false alarm rates of a blind zone alert system. Therefore, in reality, it is difficult to compare the actual performance between different systems based on either radars, cameras, or wireless technologies.

\subsection{Comparison with solutions based on vehicular ad hoc networks}
The studies in vehicular ad hoc network focus on the inter-vehicular communications~\cite{yousefi2006vehicular}\cite{chen2005ad}. In a vehicular ad hoc network, every car is equipped with a wireless device designed for the highly mobile environment, e.g., Dedicated Short Range Communications (DSRC) radio. Another possible design of the blind zone detection system is to utilize the vehicular ad hoc network~\cite{tatchikou2006cooperative}. However, since there is only one radio on each vehicle, it is not possible to rely on solely beacon packets and signal strength to determine if the target vehicle is in the blind zone. This approach might need to rely on interchanging GPS coordinates of the adjacent vehicles, and the GPS coordinates have to be very precise (which is difficult in reality). Furthermore, a DSRC radio has a much longer range of up to 200 meters, whereas the blind zone detection only considers a range less than 10 meters. Broadcasting beacons for blind zone detection would potentially cause unnecessary loads and interference to the vehicular ad hoc network. Therefore, leveraging vehicular ad hoc networks for blind zone detection might not be an efficient approach.

\subsection{Battery life estimation}
The battery life is an important concern about wireless sensors. Regarding the BLE sensor platform used in this paper, the average current consumption of sending an 8-byte beacon packet is around 7.599 mA, and the average duration is 1.028 ms~\cite{TI_AN092}. The current consumption during the sleep state is 0.9 $\mu$A. If the period of the beacon packets is 0.25 second (as assumed in Section II), the average current consumption $I_c$ can be calculated as:
\begin{align}
	I_c &= \frac{I_t T_t + I_s T_s}{T_t+T_s}\nonumber\\
	&= \frac{7.599\times 1.028+ 0.0009\times(250-1.028))}{250}\nonumber\\
	&\cong 0.032\mbox{ mA}
\end{align}
Assuming that the sensor is powered by a 3-volt lithium ion battery with capacity $C_b$ around 1000 mAh (e.g., a Panasonic BR-2477A battery), the estimated battery life $T_b$ would be:
\begin{align}
T_b =& C_b/I_c\nonumber\\ 
	=& 1000\mbox{ mAh} / 0.032\mbox{ mA}\nonumber \\
	=& 31250 \mbox{ hours} \cong 1302 \mbox{ days} \cong 3.57 \mbox{ years}
\end{align}
Moreover, the sensors only consume power when the engine is on. If assuming that, on average, the car operates 8 hours per day, the estimated battery life could be:
\begin{align}
T_b'=& T_b/D_c\nonumber\\
=& \frac{3.57 \mbox{ years}}{\frac{8}{24}} = 10.71 \mbox{ years}
\end{align}
where $D_c$ is the average duty cycle of the car. Therefore, if only the power consumption for beacon packets is considered, the beacon period is 0.25 second, and the car is driven 8 hours per day on average, then the estimated battery life of the BLE sensor node can be up to 10.71 years. It is worth pointing out that the actual battery life may be shorter due to other factors such as the temperature variation in the environment. However, the estimated battery life would still be longer than 5 years, which is the desired battery life for most vehicular applications.

\subsection{Other related issues}
The proposed system in this paper relies on communications among wireless sensors on different vehicles. The idea of near-range inter-vehicle communications via wireless sensors behind the proposed system could be applied to other safety applications in addition to the blind zone alert system. Currently, most of the sensors used by the vehicles in the market, e.g., TPMS sensors, are not fully standardized. Every car manufacturer and vendor may thus choose different and/or proprietary radio systems and protocols for wireless sensors. Standardization of wireless sensors would enable the inter-vehicle communications among those sensors and accelerate the development of related applications~\cite{nabi2011towards}. Furthermore, for the detection system proposed in this paper, minimizing the discrepancy among the systems on different vehicles, such as positions of the front and rear sensors and antenna patterns, will guarantee the system performance reported in this paper. This could be achieved through government regulations because of the potential benefits and enhanced safety. Moreover, in the future, the concept of Internet of Things (IoT)~\cite{xu2014internet} might be applied to automotive sensors, and it could lead to an open architecture for automotive electronics.

Another important concern is how to choose the beacon packet period in the proposed system and its implications. If a smaller period is used, the response time of the system, i.e., the time needed to detect a vehicle in the blind zone, will be shorter; however, the battery life of the sensors will decrease. For example, consider an overtaking scenario: the target vehicle is approaching and overtaking the detector vehicle. Assume the relative speed of the two vehicles is $v$ m/s, the length of the blind zone is $l$ meters, and the beacon period is $T$ seconds. The time for the target vehicle to travel through the blind zone of the detection vehicle would be $\frac{l}{v}$ seconds. If the computation time is disregarded and at least three beacon packets are needed for detection, the beacon period $T$ should be
\begin{align}
T \leq & \frac{l\rho}{3v}
\end{align}  
where $\rho$ represents the estimated average packet delivery ratio when the target car is in the blind zone, and it would be around 99\% as observed in the experiments. If there is strong interference expected in the environment, $\rho$ can be set as a smaller value. Note that a blind zone alert system usually does not need to detect a vehicle passing through at a high relative speed, so the range of $v$ is limited. For example, $v\leq5$ m/s is probably a reasonable assumption. Also, the defined blind zone of the system is usually larger than the actual blind zone to the driver, so the additional part can be considered as the safety cushion or guard zone. A slight delay in the detection is acceptable because of the guard zone.

\subsection{Possible enhancements and future work}
As future work, the proposed system in this paper can be improved and enhanced in several different aspects. For instance, the transmission power is an important setting that affects the system performance significantly. Under normal circumstances, a relatively low transmission power, e.g., -30 dBm, is sufficient for detecting the presence of a car in the blind zone. It is, therefore, possible to reduce the transmission power in order to extend the battery life of the sensors and reduce the chance of collisions between beacon packets from multiple vehicles in near proximity. On the other hand, when there is interference from other wireless devices operating in the same frequency band, the system performance would decline if the transmission power is too small. Further research is needed for determining the optimal transmission power. Moreover, some other techniques such as performing carrier-sense before transmitting beacons and transmitting beacons on multiple subchannels might be helpful to mitigate the interference issue. 

The proposed system utilizes the signal strength as the major feature for the detection algorithm to minimize the cost and system complexity. Also, only the signal strength of valid packets is considered by the detection system, and this mechanism effectively reduces the false alarms caused by other objects within the blind zone. In addition to signal strength, one may consider to use other metrics, such as Link Quality Indicator (LQI), packet delay, and packet round-trip time. However, packet delay might not be a good choice mainly due to synchronization issues among sensors, and packet round-trip time might not be consistent since the sensors might have other tasks that lead to varied processing delays. It would be interesting to compare the designs based on different metrics.

In addition, the training of the system can be performed separately in different driving scenarios, e.g., highways, urban areas, rural areas, various weather conditions, and etc. This way, a separate decision map is derived for each scenario. When performing the detection, the detection device first determines the current driving scenario based on speed or even map information collected from other systems in the vehicle and uses the corresponding decision map to make the decision. All the necessary driving scenarios will be identified and the experiments for those scenarios will be conducted in a future study. 

In Experiment 1 (i.e., parking scenario), the target car could be determined precisely whether it is in the blind zone based on the location measurements in the parking lot. However, In Experiment 2 (i.e., driving scenario), since both cars are moving, it is difficult to determine if the target car is actually in the blind zone with high precision. To solve this problem, a commercial radar-based system was used as the ground truth to train and test the proposed system. Nevertheless, as mentioned before, there is no perfect system on the market that could perform the detection without any error, and the choice of ground truth also affects the evaluation results of the proposed system. Instead of using a commercial system as the ground truth, a special system could be designed in order to precisely evaluate the performance of the blind zone alert system. The range of the blind zone also has to be precisely specified. For example, as future work, a fixed camera could be set up at the end of the vehicle with the predefined range of blind zone overlapped on the screen. A person could watch the camera output to decide in real time if the car is actually in the blind zone (i.e., as the ground truth).

\section{Conclusion}
It is envisioned that modern vehicles produced in the near future will be equipped with more wireless sensors, which are parts of the IVWSNs, to improve fuel economy, safety, engine performance as well as offer more features. Based on the observation that some of these wireless sensors have low duty cycles, it is shown that additional features and functionalities can be provided by utilizing the idle time of these sensors. In this paper, the blind zone alert system is chosen as an illustrative application; a specially designed system is installed in the rear of a vehicle and it detects the presence of a target vehicle in its blind zone based on the received signal strength of packets broadcast by the sensors such as TPMS sensors of the target vehicle. The system is designed, implemented, and evaluated on a commercially available BLE platform. Evaluation results from the two real experiments conducted are very promising as the proposed system can achieve approximately $95\%$ to $99\%$ detection rate with less than $15\%$ false alarm rate. Due to its low cost (as compared to the existing systems such as radar- and camera-based solutions) and simple implementation (i.e., the proposed system can be implemented on the existing sensors with slight modifications), the IVWSN-based blind zone alert system presented in this paper could be an attractive solution for car manufacturers.

\ifCLASSOPTIONcaptionsoff
  \newpage
\fi

\setstretch{0.92}
\bibliographystyle{IEEEtran}
{\tiny
\bibliography{ALL_TII-14-0330}

\begin{thebibliography}{10}
\providecommand{\url}[1]{#1}
\csname url@samestyle\endcsname
\providecommand{\newblock}{\relax}
\providecommand{\bibinfo}[2]{#2}
\providecommand{\BIBentrySTDinterwordspacing}{\spaceskip=0pt\relax}
\providecommand{\BIBentryALTinterwordstretchfactor}{4}
\providecommand{\BIBentryALTinterwordspacing}{\spaceskip=\fontdimen2\font plus
\BIBentryALTinterwordstretchfactor\fontdimen3\font minus
  \fontdimen4\font\relax}
\providecommand{\BIBforeignlanguage}[2]{{%
\expandafter\ifx\csname l@#1\endcsname\relax
\typeout{** WARNING: IEEEtran.bst: No hyphenation pattern has been}%
\typeout{** loaded for the language `#1'. Using the pattern for}%
\typeout{** the default language instead.}%
\else
\language=\csname l@#1\endcsname
\fi
#2}}
\providecommand{\BIBdecl}{\relax}
\BIBdecl

\bibitem{GM-intra-vehicular}
M.~Ahmed, C.~Saraydar, T.~ElBatt, J.~Yin, T.~Talty, and M.~Ames,
  ``Intra-vehicular wireless networks,'' in \emph{Proc. of the IEEE Globecom
  Workshops}, November 2007, pp. 1--9.

\bibitem{yick2008wireless}
J.~Yick, B.~Mukherjee, and D.~Ghosal, ``Wireless sensor network survey,''
  \emph{Computer networks}, vol.~52, no.~12, pp. 2292--2330, 2008.

\bibitem{Kong2008}
F.~Kong and J.~Tan, ``A collaboration-based hybrid vehicular sensor network
  architecture,'' in \emph{Proc. of the IEEE International Conference on
  Information and Automation (ICIA)}, 2008, pp. 584--589.

\bibitem{demirkol2006mac}
I.~Demirkol, C.~Ersoy, and F.~Alagoz, ``Mac protocols for wireless sensor
  networks: a survey,'' \emph{IEEE Communications Magazine}, vol.~44, no.~4,
  pp. 115--121, 2006.

\bibitem{ergen2010tdma}
S.~C. Ergen and P.~Varaiya, ``Tdma scheduling algorithms for wireless sensor
  networks,'' \emph{Wireless Networks}, vol.~16, no.~4, pp. 985--997, 2010.

\bibitem{ergen2006pedamacs}
------, ``Pedamacs: Power efficient and delay aware medium access protocol for
  sensor networks,'' \emph{IEEE Transactions on Mobile Computing}, vol.~5,
  no.~7, pp. 920--930, 2006.

\bibitem{shen2014prioritymac}
W.~Shen, T.~Zhang, F.~Barac, and M.~Gidlund, ``Prioritymac: A priority-enhanced
  mac protocol for critical traffic in industrial wireless sensor and actuator
  networks,'' \emph{IEEE Transactions on Industrial Informatics}, 2014.

\bibitem{toscano2012multichannel}
E.~Toscano and L.~Lo~Bello, ``Multichannel superframe scheduling for ieee
  802.15. 4 industrial wireless sensor networks,'' \emph{IEEE Transactions on
  Industrial Informatics}, vol.~8, no.~2, pp. 337--350, 2012.

\bibitem{akkaya2005survey}
K.~Akkaya and M.~Younis, ``A survey on routing protocols for wireless sensor
  networks,'' \emph{Ad hoc networks}, vol.~3, no.~3, pp. 325--349, 2005.

\bibitem{ergen2007energy}
S.~C. Ergen and P.~Varaiya, ``Energy efficient routing with delay guarantee for
  sensor networks,'' \emph{Wireless Networks}, vol.~13, no.~5, pp. 679--690,
  2007.

\bibitem{zhang2014energy}
D.~Zhang, G.~Li, K.~Zheng, X.~Ming, and Z.-H. Pan, ``An energy-balanced routing
  method based on forward-aware factor for wireless sensor networks,''
  \emph{IEEE Transactions on Industrial Informatics}, vol.~10, no.~1, pp.
  766--773, 2014.

\bibitem{niu2014r3e}
J.~Niu, L.~Cheng, Y.~Gu, L.~Shu, and S.~K. Das, ``R3e: Reliable reactive
  routing enhancement for wireless sensor networks,'' \emph{IEEE Transactions
  on Industrial Informatics}, vol.~10, no.~1, pp. 784--794, 2014.

\bibitem{bonivento2007system}
A.~Bonivento, C.~Fischione, L.~Necchi, F.~Pianegiani, and
  A.~Sangiovanni-Vincentelli, ``System level design for clustered wireless
  sensor networks,'' \emph{IEEE Transactions on Industrial Informatics},
  vol.~3, no.~3, pp. 202--214, 2007.

\bibitem{shin2010experimental}
S.~Shin, T.~Kwon, G.-Y. Jo, Y.~Park, and H.~Rhy, ``An experimental study of
  hierarchical intrusion detection for wireless industrial sensor networks,''
  \emph{IEEE Transactions on Industrial Informatics}, vol.~6, no.~4, pp.
  744--757, 2010.

\bibitem{anastasi2009energy}
G.~Anastasi, M.~Conti, M.~Di~Francesco, and A.~Passarella, ``Energy
  conservation in wireless sensor networks: A survey,'' \emph{Ad Hoc Networks},
  vol.~7, no.~3, pp. 537--568, 2009.

\bibitem{anastasi2009extending}
G.~Anastasi, M.~Conti, and M.~Di~Francesco, ``Extending the lifetime of
  wireless sensor networks through adaptive sleep,'' \emph{IEEE Transactions on
  Industrial Informatics}, vol.~5, no.~3, pp. 351--365, 2009.

\bibitem{magno2014ensuring}
M.~Magno, D.~Boyle, D.~Brunelli, E.~Popovici, and L.~Benini, ``Ensuring
  survivability of resource-intensive sensor networks through ultra-low power
  overlays,'' \emph{IEEE Transactions on Industrial Informatics}, vol.~10,
  no.~2, pp. 946--956, 2014.

\bibitem{li2014qos}
L.~Li, S.~Li, and S.~Zhao, ``Qos-aware scheduling of services-oriented internet
  of things,'' \emph{IEEE Transactions on Industrial Informatics}, vol.~10,
  no.~2, pp. 1497--1505, 2014.

\bibitem{marchenko2014experimental}
N.~Marchenko, T.~Andre, G.~Brandner, W.~Masood, and C.~Bettstetter, ``An
  experimental study of selective cooperative relaying in industrial wireless
  sensor networks,'' \emph{IEEE Transactions on Industrial Informatics},
  vol.~10, no.~3, pp. 1806--1816, 2014.

\bibitem{park2014robust}
K.~Park, J.~Kim, H.~Lim, and Y.~Eun, ``Robust path diversity for network
  quality of service in cyber-physical systems,'' \emph{IEEE Transactions on
  Industrial Informatics}, vol.~10, no.~4, pp. 2204--2215, 2014.

\bibitem{intra-car-channel}
A.~R. Moghimi, H.-M. Tsai, C.~Saraydar, and O.~K. Tonguz, ``Characterizing
  intra-car wireless channels,'' \emph{IEEE Transactions on Vehicular
  Technology}, vol.~58, no.~9, pp. 5299--5305, November 2009.

\bibitem{Lin_Globecom2013}
J.-R. Lin, T.~Talty, and O.~K. Tonguz, ``An empirical performance study of
  intra-vehicular wireless sensor networks under wifi and bluetooth
  interference,'' in \emph{Proc. of the IEEE Global Communications Conference
  (GLOBECOM)}, December 2013, pp. 581--586.

\bibitem{lee2009security}
H.~Lee, H.-M. Tsai, and O.~K. Tonguz, ``On the security of intra-car wireless
  sensor networks,'' in \emph{Proc. of the IEEE 70th Vehicular Technology
  Conference (VTC)}, 2009, pp. 1--5.

\bibitem{Tsai07}
H.-M. Tsai, O.~K. Tonguz, C.~Saraydar, T.~Talty, M.~Ames, and A.~Macdonald,
  ``Zigbee-based intra-car wireless sensor networks: A case study,'' \emph{IEEE
  Wireless Communications Magazine}, vol.~14, no.~6, pp. 67--77, December 2007.

\bibitem{BLELin14}
J.-R. Lin, T.~Talty, and O.~K. Tonguz, ``On the potential of bluetooth low
  energy technology for vehicular applications,'' \emph{IEEE Communications
  Magazine, Ad Hoc and Sensor Networks Series}, vol.~53, no.~1, pp. 267--275,
  January 2015.

\bibitem{Tsai06}
O.~K. Tonguz, H.-M. Tsai, T.~Talty, A.~Macdonald, and C.~Saraydar, ``{RFID}
  technology for intra-car communications: A new paradigm,'' in \emph{Proc. of
  the IEEE 64th Vehicular Technology Conference (VTC)}, 2006, pp. 1--6.

\bibitem{niu2008intra}
W.~Niu, J.~Li, and T.~Talty, ``Intra-vehicle uwb channel measurements and
  statistical analysis,'' in \emph{Proc. of the IEEE Global Communications
  Conference (GLOBECOM)}, 2008, pp. 1--5.

\bibitem{bas2012ultra}
C.~U. Bas and S.~C. Ergen, ``Ultra-wideband channel model for intra-vehicular
  wireless sensor networks,'' in \emph{Proc. of the IEEE Wireless
  Communications and Networking Conference (WCNC)}, 2012, pp. 42--47.

\bibitem{bas2013ultra}
------, ``Ultra-wideband channel model for intra-vehicular wireless sensor
  networks beneath the chassis: From statistical model to simulations,''
  \emph{IEEE Transactions on Vehicular Technology}, vol.~62, pp. 14--25, 2013.

\bibitem{carmo20102}
J.~P. Carmo, P.~M. Mendes, C.~Couto, and J.~H. Correia, ``A 2.4-ghz cmos
  short-range wireless-sensor-network interface for automotive applications,''
  \emph{IEEE Transactions on Industrial Electronics}, vol.~57, no.~5, pp.
  1764--1771, 2010.

\bibitem{de2009coexistence}
R.~de~Francisco, L.~Huang, G.~Dolmans, and H.~de~Groot, ``Coexistence of zigbee
  wireless sensor networks and bluetooth inside a vehicle,'' in \emph{Proc. of
  the IEEE 20th International Symposium on Personal, Indoor and Mobile Radio
  Communications}, 2009, pp. 2700--2704.

\bibitem{lo2009coexistence}
L.~Lo~Bello and E.~Toscano, ``Coexistence issues of multiple co-located ieee
  802.15.4/zigbee networks running on adjacent radio channels in industrial
  environments,'' \emph{IEEE Transactions on Industrial Informatics}, vol.~5,
  no.~2, pp. 157--167, 2009.

\bibitem{matolak2008aircraft}
D.~W. Matolak and A.~Chandrasekaran, ``Aircraft intra-vehicular channel
  characterization in the 5 ghz band,'' in \emph{Proc. of the IEEE Integrated
  Communications, Navigation and Surveillance Conference (ICNS)}, 2008, pp.
  1--6.

\bibitem{NHTSA_Plan_2013}
\BIBentryALTinterwordspacing
{National Highway Traffic Safety Administration (NHTSA)}, ``Vehicle safety and
  fuel economy rulemaking and research priority plan 2011-2013,'' 2011.
  [Online]. Available: \url{http://www.nhtsa.gov/}
\BIBentrySTDinterwordspacing

\bibitem{2011EscaladeManual}
{General Motors Inc.}, ``2011 {Cadillac Escalade} owner's manual,'' 2010.

\bibitem{TPMS_NHTSA}
{National Highway Traffic Safety Administration (NHTSA)}, ``Tire pressure
  monitoring systems,'' 2002, {Federal} Motor Vehicle Safety Standards Part
  571, Standard No. 138.

\bibitem{BLEspec}
\BIBentryALTinterwordspacing
{Bluetooth Special Interest Group}, ``{Bluetooth Core Version 4.0}
  specification,'' June 2010. [Online]. Available:
  \url{https://www.bluetooth.org/Technical/Specifications/adopted.htm}
\BIBentrySTDinterwordspacing

\bibitem{CC2540kit}
\BIBentryALTinterwordspacing
{Texas Instruments}, ``{CC2540} mini development kit,'' 2010. [Online].
  Available: \url{http://www.ti.com/tool/cc2540dk-mini}
\BIBentrySTDinterwordspacing

\bibitem{JRVTC2011}
J.-R. Lin, T.~Talty, and O.~K. Tonguz, ``Feasibility of safety applications
  based on intra-car wireless sensor networks: A case study,'' in \emph{Proc.
  of the IEEE 46th Vehicular Technology Conference (VTC)}, September 2011, pp.
  1--5.

\bibitem{RITABSWlesson}
``Design blind spot warning systems to minimize false alarms,'' {U.S.
  Department of Transportation, Research and Innovative Technology
  Administration, Knowledge Resources - Lessons Learned, January 2014}.

\bibitem{sotelo2007vision}
M.~Sotelo, J.~Barriga, D.~Fern{\'a}ndez, I.~Parra, J.~E. Naranjo,
  M.~Marr{\'o}n, S.~Alvarez, and M.~Gavil{\'a}n, ``Vision-based blind spot
  detection using optical flow,'' in \emph{Computer Aided Systems
  Theory--EUROCAST 2007}.\hskip 1em plus 0.5em minus 0.4em\relax Springer,
  2007, pp. 1113--1118.

\bibitem{alonso2008lane}
J.~D. Alonso, E.~Ros~Vidal, A.~Rotter, and M.~Muhlenberg, ``Lane-change
  decision aid system based on motion-driven vehicle tracking,'' \emph{IEEE
  Transactions on Vehicular Technology}, vol.~57, no.~5, pp. 2736--2746, 2008.

\bibitem{Lee2013nhtsa}
H.~G. Lee and S.~M. Yoo, ``Evaluation of the vision based blind spot detection
  system misjudgement performance based on roadway curvature,'' in \emph{Proc.
  of the NHTSA ESV 23rd Conference}, 2013, paper number 13-0331-W.

\bibitem{yousefi2006vehicular}
S.~Yousefi, M.~S. Mousavi, and M.~Fathy, ``Vehicular ad hoc networks (vanets):
  challenges and perspectives,'' in \emph{Proc. of the 6th IEEE International
  Conference on ITS Telecommunications}, 2006, pp. 761--766.

\bibitem{chen2005ad}
W.~Chen and S.~Cai, ``Ad hoc peer-to-peer network architecture for vehicle
  safety communications,'' \emph{IEEE Communications Magazine}, vol.~43, no.~4,
  pp. 100--107, 2005.

\bibitem{tatchikou2006cooperative}
R.~Tatchikou, S.~Biswas, and F.~Dion, ``{Cooperative vehicle collision
  avoidance using inter-vehicle packet forwarding},'' in \emph{Proc. of the
  IEEE Global Communications Conference (GLOBECOM)}, vol.~5, 2005, pp. 5--2766.

\bibitem{TI_AN092}
{Texas Instruments}, ``Measuring bluetooth low energy power consumption,''
  Application Note AN092, August 2010.

\bibitem{nabi2011towards}
Z.~Nabi, A.~Alvi, and R.~Mehmood, ``Towards standardization of in-car
  sensors,'' in \emph{Communication Technologies for Vehicles}.\hskip 1em plus
  0.5em minus 0.4em\relax Springer, 2011, pp. 216--223.

\bibitem{xu2014internet}
L.~Da~Xu, W.~He, and S.~Li, ``Internet of things in industries: A survey,''
  \emph{IEEE Transactions on Industrial Informatics}, vol.~10, no.~4, pp.
  2233--2243, 2014.

\end{thebibliography}
}

\begin{IEEEbiography}[{\includegraphics[width=1in,height=1.25in,clip,keepaspectratio]{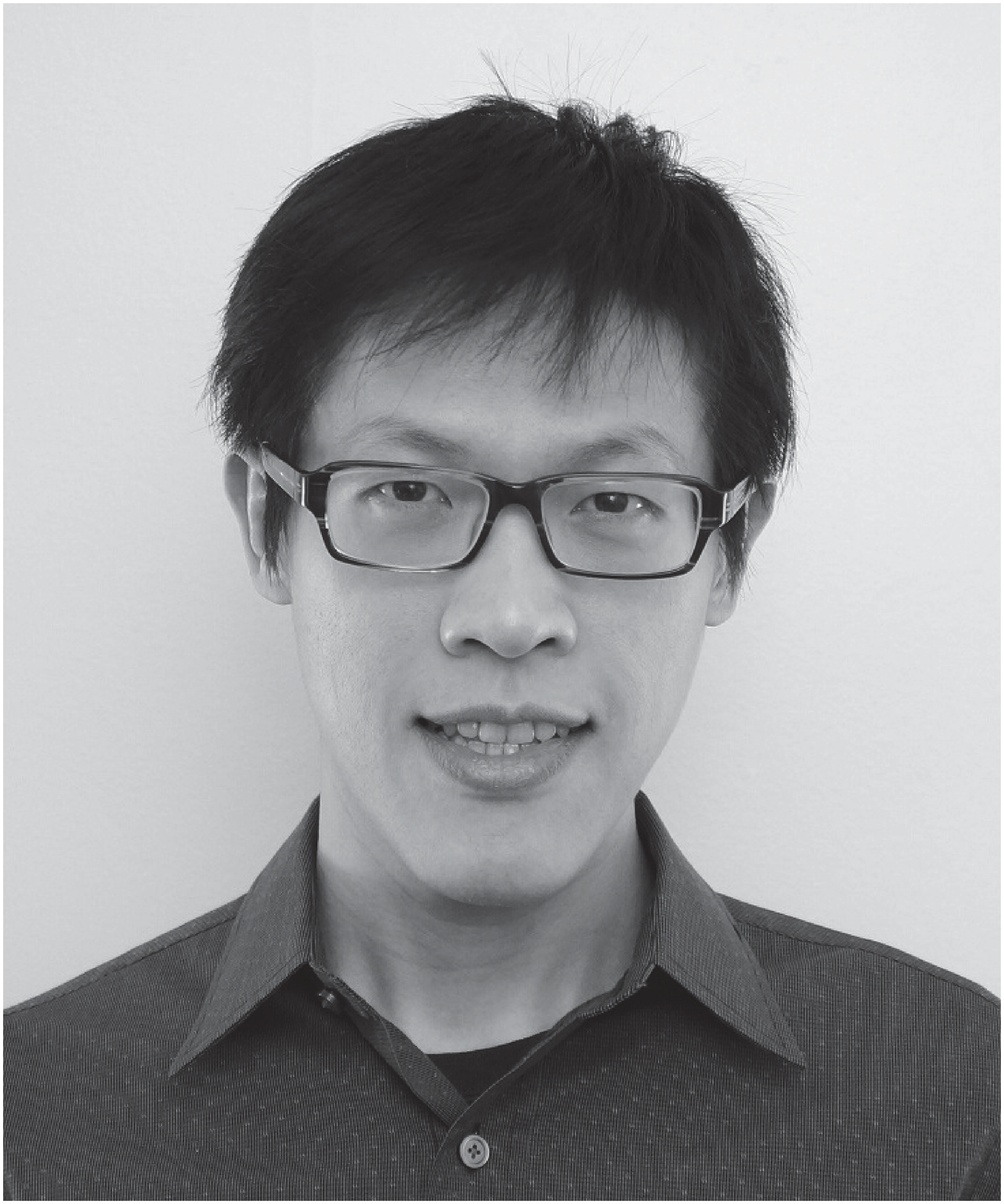}}]{Jiun-Ren Lin} (S'11)
received the B.S. degree from National Chiao Tung University, Hsinchu City, Taiwan, in 2004; the M.S. degree from National Taiwan University, Taipei City, Taiwan, in 2006, both in computer science and information engineering; and the Ph.D. degree in electrical and computer engineering from Carnegie Mellon University, Pittsburgh, PA, USA, in 2014. 

His research interests include computer networks, wireless networks and communications, personal communication systems, vehicular networks, algorithm design, performance evaluation, and machine learning. 

Dr. Lin is a member of IEEE and Eta Kappa Nu. He has served as the peer reviewer of several IEEE journals and conferences, including IEEE \textsc{Transactions on Mobile Computing}, IEEE \textsc{Wireless Communications Letters}, IEEE Global Communications Conference (GLOBECOM), IEEE Vehicular Networking Conference (VNC), IEEE Wireless Communications and Networking Conference (WCNC), IEEE International Conference on Sensing, Communications, and Networking (SECON), IEEE International Conference on Communications (ICC), IEEE Vehicular Technology Conference (VTC), and IEEE/IFIP Annual Conference on Wireless On-demand Network Systems and Services (WONS).
\end{IEEEbiography}

\begin{IEEEbiography}[{\includegraphics[width=1in,height=1.25in,clip,keepaspectratio]{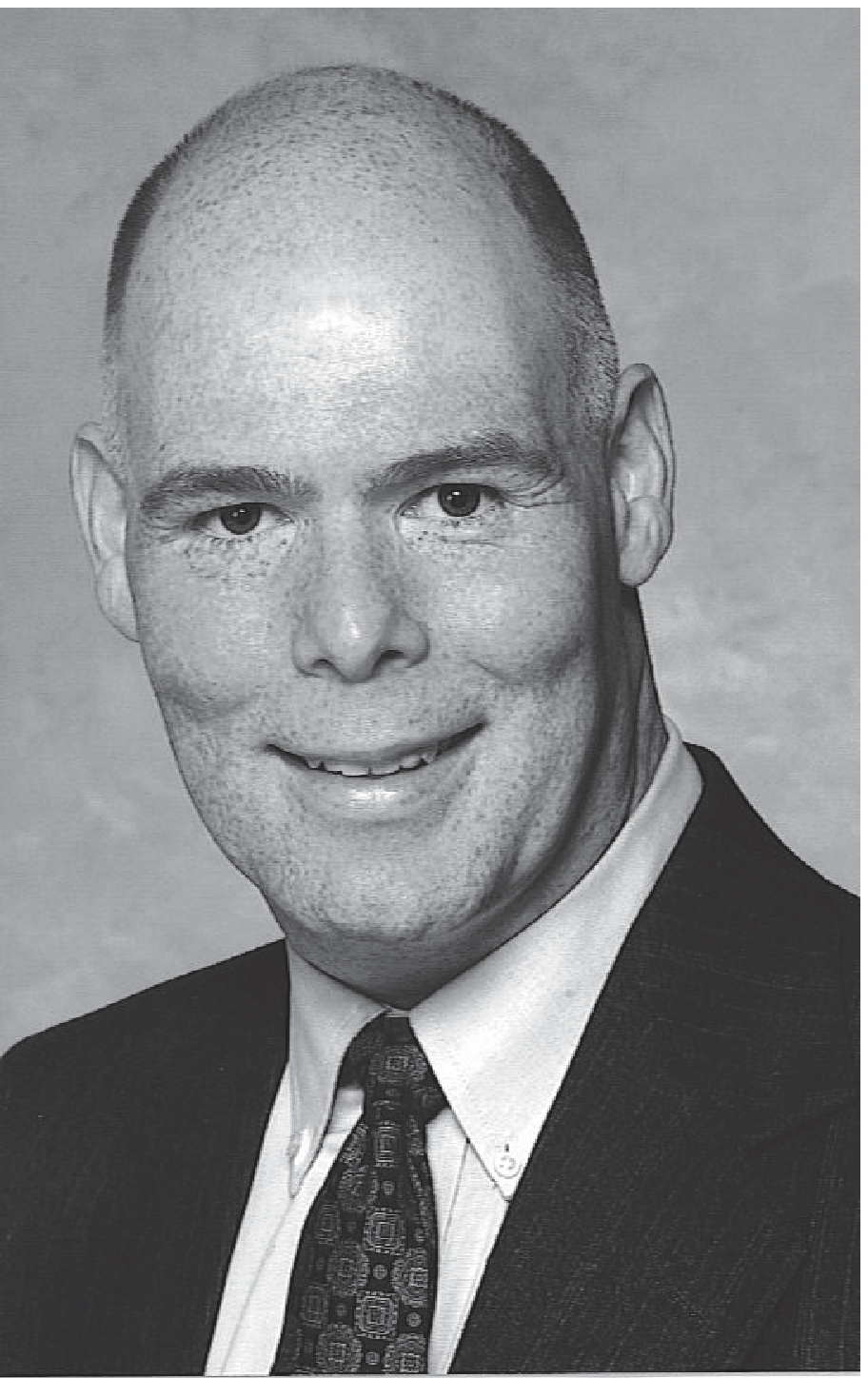}}]{Timothy Talty} (SM'85)
received the B.S. degree in electrical engineering from Trine University, Angola, IN, USA, in 1987, and M.S. and Ph.D. degrees in electrical engineering from The University of Toledo, Toledo, OH, USA, in 1996, under the direction of Dr. Kai-Fong Lee.  

He is currently a Technical Fellow with General Motors (GM) Research and Development, Warren, MI, USA.  He is also an Adjunct Professor with the Electrical and Computer Engineering Department, University of Michigan--Dearborn, MI, USA.  Prior positions include Group Manager at HMI \& Infotainment with GM, Warren, MI, USA; Supervisor Radios and Antennas with Ford Motor Company, Detroit, MI, USA; and Assistant Professor with the United States Military Academy, West Point, NY, USA. In his current role as a technical fellow with GM, he provides the technical leadership of the development of advanced wireless technologies for automotive use cases. He has authored numerous papers and 21 patents. His research interests include infotainment, telematics, and wireless communication technologies.
\end{IEEEbiography}

\begin{IEEEbiography}[{\includegraphics[width=1in,height=1.25in,clip,keepaspectratio]{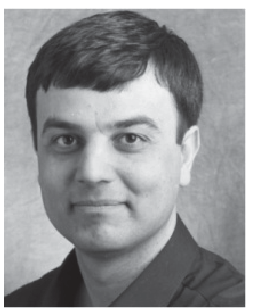}}]{Ozan K. Tonguz} (M'90) received his Ph.D. in electrical and computer engineering from Rutgers University, Piscataway, New Jersey, in 1990. He currently serves as a tenured Full Professor in the Department of Electrical and Computer Engineering at Carnegie Mellon University (CMU), Pittsburgh, PA, USA.  Before joining CMU in August 2000, he was with the Electrical and Computer Engineering Department of the State University of New York at Buffalo (SUNY/Buffalo), Buffalo, NY, USA. He joined SUNY/Buffalo in 1990 as an Assistant Professor, where he was granted early tenure and promoted to Associate Professor in 1995, and to Full Professor in 1998. 

Prior to joining academia, he was with Bell Communications Research (Bellcore) in Red Bank, NJ, USA, from 1988 to 1990 doing research in optical networks and communication systems. He has authored about 300 technical papers in IEEE journals and conference proceedings. He is well-known for his contributions to vehicular networks, wireless communications and networks, and optical communications and networks. He is the author (with G. Ferrari) of Ad Hoc Wireless Networks: A Communication-Theoretic Perspective (Wiley, 2006). He is the founder and President of Virtual Traffic Lights, LLC, a CMU start-up which was launched in Pittsburgh, PA, in December 2010, for providing solutions to several key transportation problems related to safety and traffic information systems, using vehicle-to-vehicle and vehicle-to-infrastructure communications paradigms. His industrial experience includes periods with Bell Communications Research, CTI Inc., Red Bank, NJ; Harris RF Communications, Rochester, NY; Aria Wireless Systems, Buffalo, NY; Clearwire, Buffalo, NY; Nokia Networks, Dallas, Texas; Nokia Research Center, Boston, MA; Neuro Kinetics, Pittsburgh, PA; Asea Brown Boveri (ABB), Oslo, Norway; ABB, Raleigh, NC; General Motors (GM), Detroit, MI;  Texas Instruments, Dallas, TX; and Intel, Hillsboro, OR, USA. He currently serves or has served as a consultant or expert for several companies (such as Aria Wireless Systems, Harris RF Communications, Clearwire Technologies, Nokia Networks, Alcatel, Lucent Technologies), major law firms (Jones Day, Pittsburgh, PA; WilmerHale, NYC; NY; Williams and Connolly, Washington, DC; Heller Ehrman, San Diego, CA; Baker Botts, Dallas, TX; Soroker-Agmon, TelAviv, Israel; Dinsmore\&Shohl, Pittsburgh, PA; Carlson and Caspers, Minneapolis, MN; etc.), and government agencies (such as National Science Foundation (NSF) and United States Department of Transportation (DOT)) in the USA, Europe, and Asia in the broad area of telecommunications and networking. He also served as the Co-Director (Thrust Leader) of the Center for Wireless and Broadband Networking Research at CMU. His current research interests are in vehicular networks, sensor networks, computer networks, wireless networks and communications systems, self-organizing networks, smart grid, Internet of Things, optical communications and networks, bioinformatics, and security. More details about his research interests, research group, projects, and publications can be found on his CMU homepage.

Dr. Tonguz has served as an Associate Editor or Guest Editor for several IEEE Journals and Transactions (such as IEEE \textsc{Transactions on Communications}, IEEE \textsc{Journal on Selected Areas in Communications}, IEEE/OSA \textsc{Journal of Lightwave Technology}, IEEE \textsc{Communications Magazine}) and as a member of Technical Program Committees of several IEEE conferences and symposia.

\end{IEEEbiography}

\end{document}